# The Influence of Ceramide Tail Length on the Structure of Bilayers Composed of Stratum Corneum Lipids

T. C. Moore, R. Hartkamp, C. R. Iacovella, A. L. Bunge, and C. M[c]Cabe

Condensed Title: Stratum Corneum Lipid Bilayers

## ABSTRACT

Lipid bilayers composed of non-alpha hydroxy sphingosine ceramide (CER NS), cholesterol (CHOL), and free fatty acids (FFA), which are components of the human skin barrier, are studied via molecular dynamics simulations. Since mixtures of these lipids exist in dense gel phases with little molecular mobility at physiological conditions, care must be taken to ensure that the simulations become decorrelated from the initial conditions. Thus, we propose and validate an equilibration protocol based on simulated tempering in which the simulation takes a random walk through temperature space, allowing the system to break out of metastable configurations and hence become decorrelated form its initial configuration. After validating the equilibration protocol, the effects of the lipid composition and ceramide tail length on bilayer properties are studied. Systems containing pure CER NS, CER NS + CHOL, and CER NS + CHOL + FFA, with the CER fatty acid tail length varied within each CER NS-CHOL-FFA composition, are simulated. The bilayer thickness is found to depend on the structure of the center of the bilayer, which arises as a result of the tail length asymmetry between the lipids studied. The hydrogen bonding between the lipid headgroups and with water is found to change with the overall lipid composition, but is mostly independent of the CER fatty acid tail length. Subtle differences in the lateral packing of the lipid tails are also found as a function of CER tail length. Overall, these results provide insight into the experimentally observed trend of altered barrier properties in skin systems where there are more ceramides with shorter tails present.

## INTRODUCTION

The stratum corneum (SC) layer of the skin acts as the main barrier against chemical penetrants and water loss and is composed of a brick-and-mortar-like arrangement of corneocytes surrounded by a dense, lamellar-structured extracellular lipid matrix. This matrix is a complex mixture primarily composed of an equimolar ratio of ceramides (CERs), which are composed of a sphingoid base linked to a saturated fatty acid chain (in most cases), cholesterol (CHOL), and free fatty acids (FFAs) (1, 2). Since the extracellular matrix is widely thought to be the only continuous path for permeation through the SC, proper lipid organization and composition of the lamellar membrane is generally considered crucial for a fully functioning skin barrier. This is supported by the fact that many skin diseases that are characterized by a reduced barrier function also exhibit an altered lipid composition and organization compared to healthy skin (3-6). For example, a decrease has been observed in the average number of CER carbon atoms in the SC of patients who suffer from atopic eczema (3, 7) and, compared with controls, less dense lipid packing and changes in x-ray scattering patterns (3). Of the two CER tails, reduced CER carbon numbers appear to correspond with a shorter fatty acid (FA) tail rather than the sphingoid base tail (8).

The complex nature of the human SC lipid matrix, in that it features 15 unique subclasses of CER lipids and a distribution of CER and FFA chain lengths, has made it challenging to establish the influence of individual lipid species on barrier properties (9). As a result, many experiments have focused on model systems that are composed of simpler mixtures of synthetic lipids, whose composition and chain lengths can be precisely controlled (10-21). For example, considering studies of mixtures containing non-alpha hydroxyl sphingosine CER (NS), CHOL, and FFA, similar to those considered in this work, Mojumdar *et al.* (14) showed that CHOL is essential to the characteristic phase behavior, and hence barrier properties, of such mixtures, and Školová *et al*. (19) found that CER NS-based membranes with short acyl chains had increased permeability compared to long acyl chain CERs. However, while global structural properties of SC lipid lamellae can be inferred from experiment (e.g., from spectroscopic, scattering, and diffraction data), including information about the localization of molecules across the lamellae, the underlying molecular-level details and in-plane morphology often remain elusive. Furthermore, structural examination using neutron scattering is quite laborious, as many independent systems, each with different deuterated lipids and/or concentrations of deuterated lipids, must be considered to provide a single picture of the structure, and the availability of deuterated lipids may limit the study of specific systems (15, 22-24).

Molecular simulation, which offers atomic-level resolution of the entire 3-dimensional structure of the lipids at picosecond timescales, has been used to provide a more direct understanding of the molecular interactions in CER-based membranes (25-38), with several studies focusing on the dependence of membrane structure on lipid composition. For example, Das *et al*. (26) found that CHOL compresses a CER NS bilayer membrane and increases the interdigitation of CER tails in opposing bilayer leaflets. Gupta and Rai (31) simulated a series of lipid bilayers composed of CER NS, CHOL, and FFA with varying compositions and found that CHOL tends to sit away from the lipid-water interface and increases the thermal stability of the bilayers. The effect of the CER tail length on bilayer properties has received considerably less attention. Paloncyová *et al*. showed that extremely short CER NS FA tails (e.g., 6 carbons) cause a disruption in the headgroup packing of CER bilayers, which gives rise to an increased permeability (37); however, the CER NS FA tail lengths considered were much shorter than those found in the SC. More recently, Gupta *et al*. showed that the permeability of water across pure CER NS bilayers with more biologically relevant tail lengths decreases monotonically with the CER FA tail length (30), although a structural explanation for this trend was not presented.

While these computational studies have provided important insight, they also highlight a significant challenge in studying SC lipids with molecular simulation. Specifically, due to the dense packing of the lipid tails, the CER-rich bilayer systems demonstrate negligible molecular diffusion at physiologically relevant temperatures (i.e., 305 K). Even in simulations conducted at elevated temperatures, e.g., 340 K, studies have demonstrated minimal in-plane diffusion of the lipids, despite considering simulations in excess of 200 ns (32). Additionally, lipids in the gel phase have low rates of rotational relaxation (39). As such, on typical simulation timescales, systems likely exist in metastable configurations that are highly dependent on the initial bilayer structure. While this may be of little consequence for single component lipid bilayers, many properties of multicomponent lipid mixtures depend on the in-plane morphology of the individual lipid components, e.g., aggregation versus dispersion of lipid species in the bilayer leaflets; furthermore, the morphology on this size scale is typically not known from experiment (40). While self-assembled bilayers would, in principle, avoid these problems, it is not currently practical to study the self-assembly of multicomponent bilayers with atomistically detailed models due to the high computational cost and a rough free energy landscape that could hinder the formation of equilibrium structures; for example, Das *et al.* found inverted micelle structures formed in mixtures of 3 different CERs with CHOL and FFA unless constrained by a wall (27), in contrast to experiments, where lamella structures form in bulk. Therefore, since preassembled structures remain more practical for simulating atomistic models, care must be taken that system properties are not biased by the initial configuration, so that reliable conclusions can be made.

Here, molecular dynamics (MD) simulations are used to examine the structural behavior of model SC membranes, similar to those studied experimentally (19, 41), to examine the role of the CER FA tail length and overall lipid composition on structural properties. Specifically, this work focuses on CER NS, as this is the most abundant CER species in healthy human SC and is typically used in model systems. Three sets of systems are studied: pure CER NS, binary mixtures of CER NS and CHOL in 2-1 and 1-1 molar ratios, and equimolar mixtures of CER NS, CHOL, and FFA C24:0. The structure of each of the lipids used in this work is shown in **Figure 1**. Mixtures of different length CERs are used in each set to examine the impact of the CER FA tail length on the systems, which, to date, has not been the focus of any computational studies. To ensure robust results, a random walk MD (RWMD) algorithm is validated and employed to reduce the dependence of the final bilayer configuration on the assumed, initial morphology.

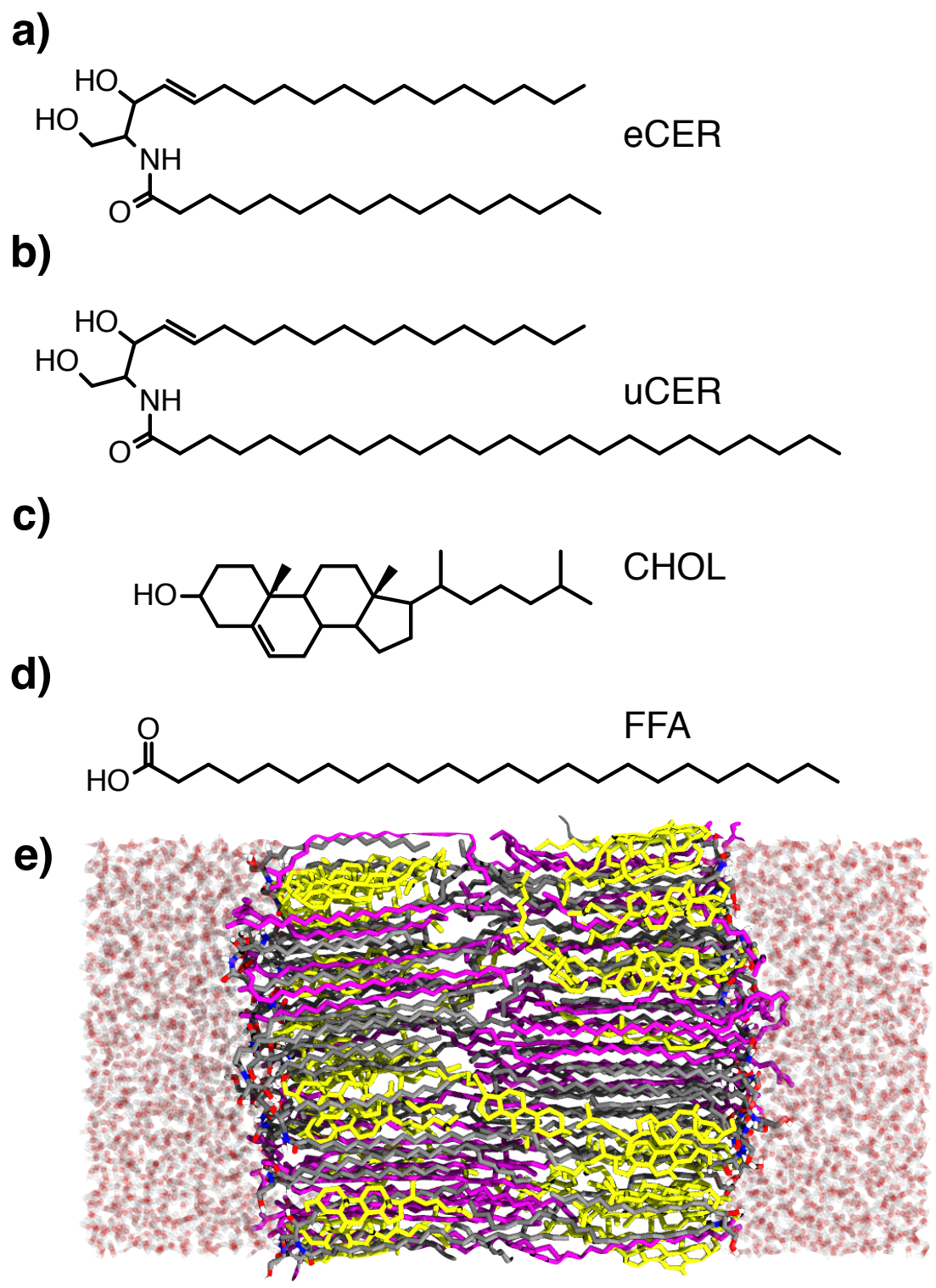

**Figure 1.** Chemical structure of the lipids in this study. a) CER NS C16 (eCER); b) CER NS C24 (uCER); c) cholesterol (CHOL); d) FFA C24:0 (FFA); and e) a snapshot of a typical configuration for the equimolar uCER2-CHOL-FFA bilayer. CER is shown in gray, CHOL is shown in yellow, FFA is shown in purple, and water is shown in transparent red (oxygens) and white (hydrogen).

## MATERIALS AND METHODS

### Model

CER NS consists of a sphingosine base, 18 carbons in length, linked to a saturated fatty acid (FA) chain of variable length. Two different FA tail lengths were considered: CER NS C16:0 (**Figure 1**a), in which the FA tail is 16 carbons long and approximately equal in length to the sphingosine chain (denoted as equal-length CER NS, or eCER); and CER NS C24:0 (**Figure 1**b), in which the FA tail is 24 carbons long, a typical length in the SC(11) (denoted as unequal-length CER NS, or uCER). FFA, when present in this work, is always 24 carbons in length (**Figure 1**d), which has been shown to be a useful model for the FFAs in healthy human SC (11, 42). The fully atomistic CHARMM36 force field (43) supplemented by the CHARMM-compatible CER headgroup parameters from Guo *et al*. (29) were used to describe the CER NS, CHOL, and FFA. The CER headgroup parameters have been shown to accurately reproduce the thermotropic phase behavior of bilayers composed of pure CER NS (29), and have been used to study the hydration of CER NS headgroups in solution (44, 45). Water was modeled using TIP3P (46).

### Methods

Lipid monolayers were first constructed by placing 64 lipids on a rectangular lattice in the *xy* plane with 50 Å$^2$ per lipid. Four initial lateral distributions of a binary mixture of eCER and CHOL (i.e., different arrangements of lipid species on the lattice) were considered for validating the ST-based approach: completely phase-separated, consisting of one 8 x 4 block of either lipid species (Figure 2a); a coarse-grained checkerboard, consisting of alternating 4 x 4 blocks of each lipid species (Figure 2b); randomly mixed, where both lipid species are randomly dispersed throughout the lattice (Figure 2c), and a fine-grained checkerboard, consisting of alternating lipid species on the lattice sites (Figure 2d). To examine the impact of the CER NS FA chain length on the structural properties, each of the four lipid compositions studied (pure CER NS, 2-1 CER NS-CHOL, 1-1 CER NS-CHOL, and equimolar CER NS-CHOL-FFA) was subdivided into five subcompositions, where the CER NS fraction consists of 0, 25, 50, 75, 100 mol% eCER. Thus, in total, 20 such systems were considered. For these simulations, the effect of the initial lateral distributions of the lipids was not explored, and the different lipid species were initially randomly dispersed throughout the lattice, with the number of each specific lipid dictated by the system composition.

In all cases, each lipid was rotated about its long axis by a random integer multiple of 60°, since a high degree of alignment between lipid backbones has been shown to cause unphysically large tilt angles in the lipid tails of gel-phase bilayers (39). The monolayer was then rotated about the $x$-axis, and translated in $z$ such that the ends of the tails in opposing leaflets were in contact, forming a bilayer; note, in this procedure each leaflet starts with the same in-plane configuration. Twenty water molecules per lipid were then added to hydrate the outside of the bilayers (2,560 total water molecules for 128 total lipids). System sizes ranged from 19,136 to 24,192 atoms.

The systems were relaxed through a series of energy minimization, NVT simulations, and NPT simulations. First, a steepest descent energy minimization was performed on each initial configuration to reduce the large non-bonded repulsions caused by (inadvertently) overlapping atoms. Next, the systems were simulated for 10 ps in the NVT ensemble at 305 K (i.e., skin temperature), followed by a 10 ns simulation in the NPT ensemble at 305 K and 1 atm. Unless otherwise noted, the systems were further simulated using the RWMD algorithm. In the RWMD algorithm, the system temperature is adjusted at small time intervals, such that the system takes a random walk through temperature space. Specifically, the sequence is defined within an interval between $T_{min}$ and $T_{max}$, with discrete temperatures defined every $\Delta T$. In this work, RWMD was performed for 50 ns, with temperature changes of $\Delta T = 5 \times j$ $(j \in \{-1, 0, 1\})$ $K$ every 5 ps, $T_{min}$ = 305 K, and $T_{max}$ set to 355 K for the first 25 ns, and the upper bound linearly reduced to 305 K over the final 25 ns. A representative plot of temperature versus time during the RWMD equilibration is shown in Figure S1, where we note the sequence is defined such that it samples all temperature states equally. Systems were further simulated at 305 K for at least 150 ns after ST. Note that 200 ns of simulation at a fixed temperature range are used for the comparison to the results obtained from simulations with RWMD in the RWMD validation section (i.e., for the RWMD validation, the 305 K systems were simulated for 200 ns at 305 K, and the RWMD systems were simulated for 200 ns with $T_{min}$ = 305 K and $T_{max}$ = 355 K). To avoid modification to the simulation code itself, we note this random walk was determined separately and the sequence of temperatures provided to the thermostat.

The RWMD approach is designed to mimic key aspects of the Monte Carlo simulated tempering (ST) algorithm (47), used to find minima on a rough free energy surface by performing a random walk through temperature space. The general idea of the ST algorithm is that increasing the system temperature lowers free energy barriers, allowing different local minima to be explored each time the system returns from an elevated temperature to the (lower) temperature of interest, all while keeping the system at equilibrium due to the short timescale (47). The latter distinguishes this approach from the widely-used simulated annealing method, in which a system is driven out of equilibrium by increasing its temperature and simulating for a long time at the elevated temperature, followed by a slow cooling to the temperature of interest.

We note that the original ST algorithm involves attempting a temperature swap after each sweep in a Monte Carlo simulation to achieve a random walk through temperature space using the standard Metropolis acceptance/rejection criteria, rather than the predetermined walk used in the RWMD implementation here.

All simulations were performed in GROMACS 5.1 (48), employing the Nosé-Hoover thermostat (49) with a coupling constant of 1 ps and, for the NPT simulations, a Parrinello-Rahman barostat (50) with a coupling constant of 10 ps. The pressure was controlled semi-isotropically for all NPT simulations, where the box lengths of the bilayer lateral directions were coupled. van der Waals interactions were smoothly switched off between 10-12 Å, beyond which they were neglected. Long-range electrostatic interactions were treated via the PME algorithm (51) with a real-space cutoff of 12 Å. A timestep of 1 fs was used for all simulations. Note that separate thermostats were used for the lipids and the water, although both groups follow the same temperature walk to avoid introducing temperature gradients into the system.

**Analysis**

System properties were calculated over the final 100 ns of simulation at 305 K. To quantify the structure of the bilayer, various properties were calculated, including: the area per lipid (*APL*), the tilt angle of the lipid tails with respect to the bilayer normal (i.e., the $z$-axis), the area per tail (*APT*), the nematic order parameter, density profiles of various groups across the bilayer normal, the bilayer thickness, the width of the lipid-water interface, the width of the low density tail region, and the hydrogen bonding. The coordination numbers (*CN*s) of different lipid tail pairs in the bilayer plane are used to describe the lateral distributions of lipids; the center of mass of each tail is projected onto the $z = 0$ plane for these calculations. Note that the tails were chosen for the *CN* calculations, as the lipid tails pack in an ordered hexagonal lattice, in contrast to the relatively more disordered packing of the lipid headgroups. The lipid backbone orientation is used to describe the rotational motion of the lipids about their long axes. A detailed description of each of these calculations is given in the Supporting Information.

## RESULTS

### Random Walk Molecular Dynamics

Before using the RWMD scheme in the rest of this work, the efficacy of the approach is first evaluated for equimolar mixtures of eCER and CHOL. This composition was chosen as eCER and CHOL have similar hydrophobic lengths, which should yield bilayers without a considerable interdigitation region in the middle; prior work has shown that CHOL can reside in this interdigitation region in lipid mixtures with uCER (28). Thus, the lipids (especially CHOL) in this mixture should stay in their canonical bilayer conformation, i.e., with their headgroups at the lipid-water interface and their tails creating a hydrophobic core, therefore allowing evaluation of the ability of RWMD to enhance the in-plane lateral and rotational rearrangements of the lipids.

**Figure 2** shows the initial configuration of each system studied, the final configurations at 200 ns at 305 K, the final configurations after 200 ns of RWMD, and the evolution of the CHOL-CHOL *CN*s for each equilibration scheme. Note that the CHOL-CHOL *CN* was chosen because it is the least biased *CN* (e.g., the FA-SPH *CN* is skewed because the FA and SPH tails are part of the same molecule, and hence will show a high level of association). Since lipids of the equimolar eCER-CHOL mixture are expected to mix, the fully separated and coarse-grained checkerboard morphologies (Figures 2a and 2b) serve as “bad” initial configurations, and the randomly mixed and fine-grained checkerboard systems (Figures 2c and 2d) serve as naïve, but reasonable, guesses of how to initialize mixed-lipid bilayers. Each initial morphology was equilibrated for 200 ns with two different schemes: standard MD at 305 K, and RWMD with 305 K < $T$ < 355 K. Both visual inspection and the evolution of the CHOL-CHOL *CN* can be used to compare the in-plane morphologies, and hence equilibration, of the different systems and equilibration procedures. At 305 K, the final morphologies visually resemble the corresponding initial morphologies; this is reflected in the steady-state nature of the *CN*s at 305 K, which are listed in Table 1. This is especially notable for the fully-separated and coarse-grained checkerboard systems, where the large CHOL aggregates make it easy to visually establish that there are only slight changes in the shape of the initial aggregates. These results highlight the frozen nature of these systems at 305 K; only small rearrangements occur in 200 ns if RWMD is not applied. Note that in the fine-grained (FG) checkerboard system, small linear aggregates of CHOL form relatively quickly, indicated by the increase in the *CN* during the first 25 ns, although the system does not evolve much after this initial change.

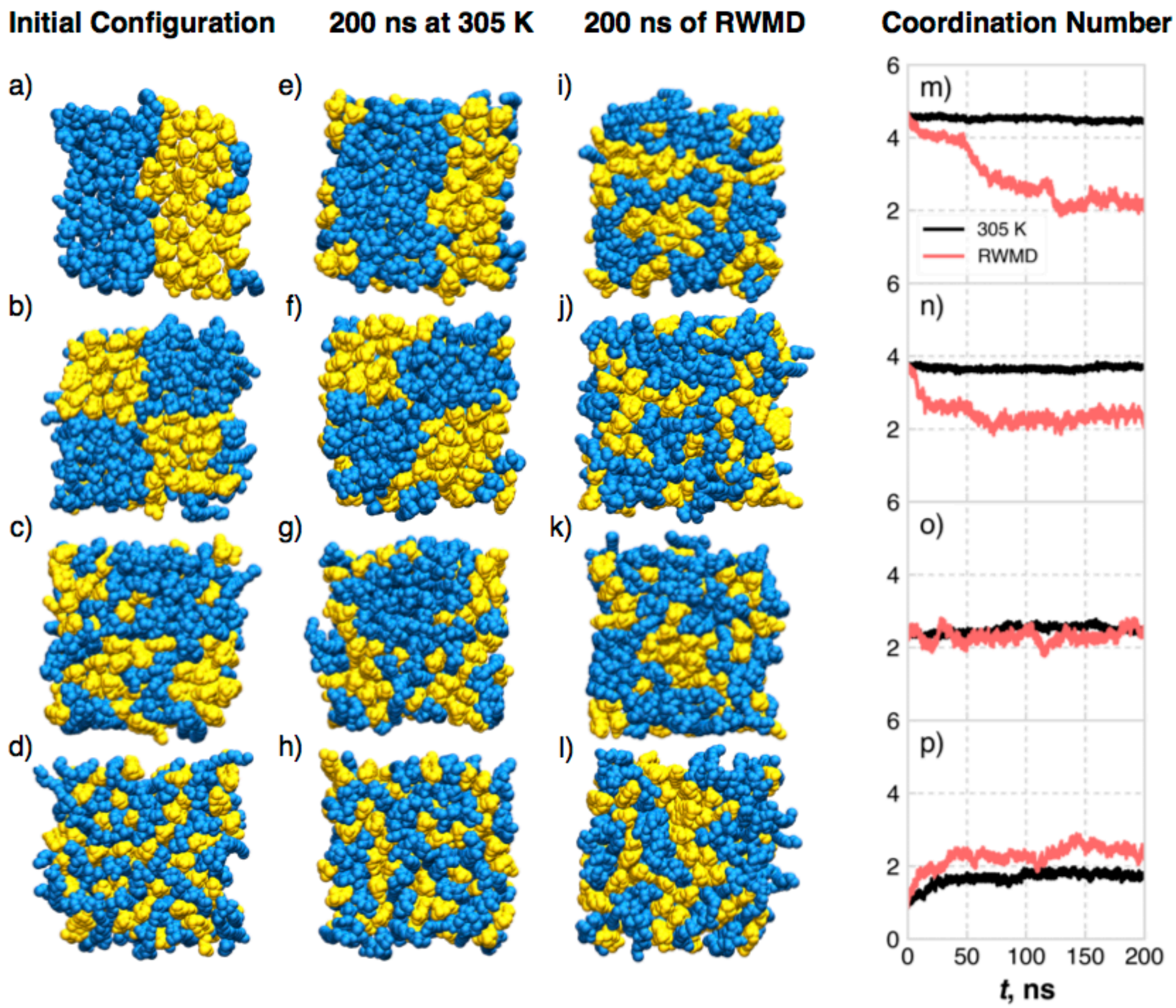

**Figure 2.** System configurations and CHOL-CHOL coordination number (*CN*) as a function of initial configuration and equilibration procedure for the equimolar mixture of eCER2 and CHOL. The first column shows the different initial configurations: a) maximum phase separation; b) coarse-grained checkerboard; c) randomly mixed; and d) fine-grained checkerboard. The second column shows the systems after 200 ns of MD at 305 K, the third column shows the systems after 200 ns of RWMD, and the fourth column shows the evolution of the CHOL-CHOL *CN* for each initial configuration. Snapshots are taken along the *z*-axis, i.e., along the bilayer normal direction. eCER2 is represented as blue spheres, CHOL is represented as yellow spheres. For clarity, water is not shown.

**Table 1.** CHOL-CHOL coordination number over the final 50 ns of MD simulation. *Separated* refers to the initial configuration shown in Figure 2a, *CG Checker* refers to that in Figure 2b, *Random Mix* refers to Figure 2c, and *FG Checker* refers to Figure 2d. The uncertainties here, and in all subsequent tables and figures, are given as the standard error of the measured property.

| | Initial Configuration | | | |
|---|---|---|---|---|
| **Equilibration** | **Separated** | **CG Checker** | **Random Mix** | **FG Checker** |
| 305 K | 4.46 ± 0.03 | 3.72 ± 0.04 | 2.53 ± 0.08 | 1.75 ± 0.07 |
| RWMD | 2.3 ± 0.1 | 2.4 ± 0.1 | 2.4 ± 0.2 | 2.5 ± 0.1 |

With RWMD, the final lateral distributions of the lipids are visually distinct from the initial configuration, showing that the lipids were able to reorganize within the bilayer leaflets. For the two most separated systems, the larger CHOL aggregates mostly break up into smaller aggregates (**Figure 2**a, i and **Figure 2**b, j), while small CHOL aggregates form in the fine-grained checkerboard system (**Figure 2**d, l). Importantly, all of the final configurations after RWMD visually resemble each other and have nearly identical *CN* values, as listed in Table 1, and systems that start close to the final morphology (i.e., the random and fine-grained checkerboard) rapidly converge to their final coordination number. Since each of these systems started from qualitatively different initial configurations, this result shows that the final configurations are decorrelated from the initial configurations and RWMD can reproducibly form the structures with matching properties without strong bias from the starting configuration. We note this does not necessarily indicate that RWMD has found the true free-energy minimum of the system, but this does strongly suggest that it is a stable, low energy configuration, given that multiple independent trajectories converge to the same state.

Analysis of the rotational motion of the eCER molecules also suggests that the systems equilibrated with RWMD are more decorrelated from their initial configurations than those equilibrated at 305 K. **Figure 3** shows the autocorrelation function of the lipid backbone angle in the bilayer plane for the different initial configurations and equilibration methods. The lipid backbone orientations become completely uncorrelated from the initial orientations by 25 ns of RWMD, while slower relaxation is observed for the systems simulated at 305 K. Interestingly, the phase-separated system shows the slowest relaxation of the eCER backbone orientations, likely because the large, dense eCER domain has a larger energy barrier for lipid rotation due to lipid-lipid hydrogen bonding. Since the systems that were equilibrated at 305 K all converge to different morphologies and the lipid backbone orientations are correlated to the initial orientations, at least three of the four systems must be metastable configurations. Often in bilayer simulations, a specific property, such as the area per lipid (*APL*), is monitored over time, and equilibrium is assumed when that property reaches a steady state. This practice, however, may lead to spurious assumptions of equilibrium since other features of the bilayer structure may not exist in their preferred state, but this cannot be easily verified since the preferred state is not always known *a priori*. For example, the *APL*s of the systems equilibrated at 305 K are compared to the *APL*s of the systems equilibrated with RWMD (after running at 305 K for a more direct comparison) in **Table 2,** from which it is apparent that the systems equilibrated at 305 K show a larger spread in the *APL*s than the systems equilibrated with RWMD. The two systems with the highest level of CHOL aggregation have similar *APL*s, while the other two systems have lower *APL*s that are similar to each other. Due to the different morphologies and packing densities of the lipids in these systems, they likely have different properties, e.g., permeability, which is related to the in-plane density of the lipids (52). In contrast, all of the systems equilibrated with RWMD have *APL*s consistent within the standard deviation, further illustrating the reproducibility, and hence confidence gained, when equilibrating with RWMD.

**Table 2.** Area per lipid (in $Å^2$) of systems as a function of initial configuration and equilibration methodology. The systems equilibrated with RWMD were simulated at 305 K for 50 ns for a more direct comparison with the other systems. *Separated* refers to Figure 2a, *CG Checker* refers to Figure 2b, *Random Mix* refers to Figure 2c, and *FG Checker* refers to Figure 2d.

| | Initial Configuration | | | |
|---|---|---|---|---|
| **Equilibration** | **Separated** | **CG Checker** | **Random Mix** | **FG Checker** |
| 305 K | 38.8 ± 0.4 | 38.8 ± 0.4 | 39.4 ± 0.2 | 39.6 ± 0.2 |
| RWMD | 39.6 ± 0.5 | 39.7 ± 0.4 | 39.5 ± 0.4 | 39.4 ± 0.3 |

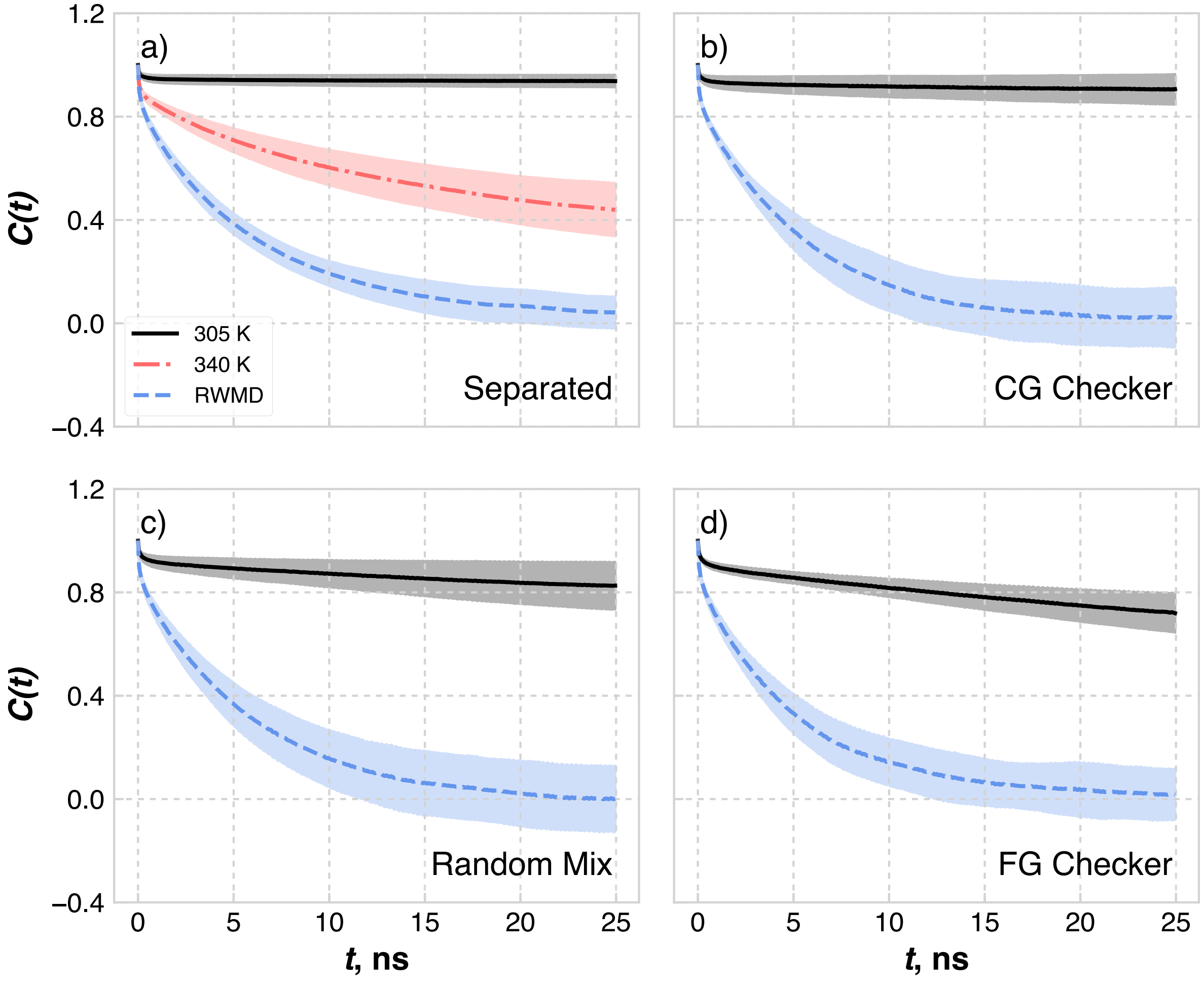

**Figure 3.** Autocorrelation function *C(t)* of the eCER backbone orientations during simulations at 305 K and with RWMD from different initial configurations. *Separated* refers to Figure 2a, *CG Checker* refers to Figure 2b, *Random Mix* refers to Figure 2c, and *FG Checker* refers to Figure 2d. The shaded area shown is the standard deviation over the rotational autocorrelation function of all eCER molecules in the bilayer.

This lack of mobility and difficulty in proper equilibration is a recognized problem when simulating CER-based bilayers, with many studies running the systems at elevated temperatures in an attempt to overcome mobility issues (25-28, 30, 31, 34, 35). Here, we also compare to equilibration and production runs at 340 K, as several MD studies of SC lipids have used this temperature (25-28). **Figure 4** shows the evolution of the *CN* for the fully separated system (Fig. 2a) simulated for 600 ns at 305 K, 340 K, and with RWMD. As expected, the system at 305 K is still *frozen* in simulations in excess of 0.5 μs, while the system simulated with RWMD reaches its steady state *CN* number after ~110 ns. Perhaps unsurprisingly, the system simulated at 340 K shows behavior somewhere in between that of the other systems; while lateral reorganization does occur, the *CN* reaches a steady state value after ~520 ns, which we note is significantly longer than the simulation times typically studied. Note that this system also reaches a pseudo-steady state *CN* between 80 and 150 ns; with many studies performed on simulations run for 100 ns or less, this metastable state could be erroneously mistaken for equilibrium, if one were relying on *CN* to determine convergence. At 600 ns, the rotational relaxation of the lipids for this system at 340 K is higher than 305 K, but still lower than RWMD (**Figure 3**a). While the fully phase-separated system is likely a poorly chosen starting configuration given some level of prior knowledge of this particular system, we emphasize that the actual morphology is typically not known *a priori*. Thus, these results indicate that RWMD is more efficient than simply running at a high temperature to equilibrate multicomponent, gel-phase lipid bilayers and can provide increased confidence in the reproducibility of the results, particularly for systems in which the lateral organization is unknown and expected to play an important role in properties of interest.

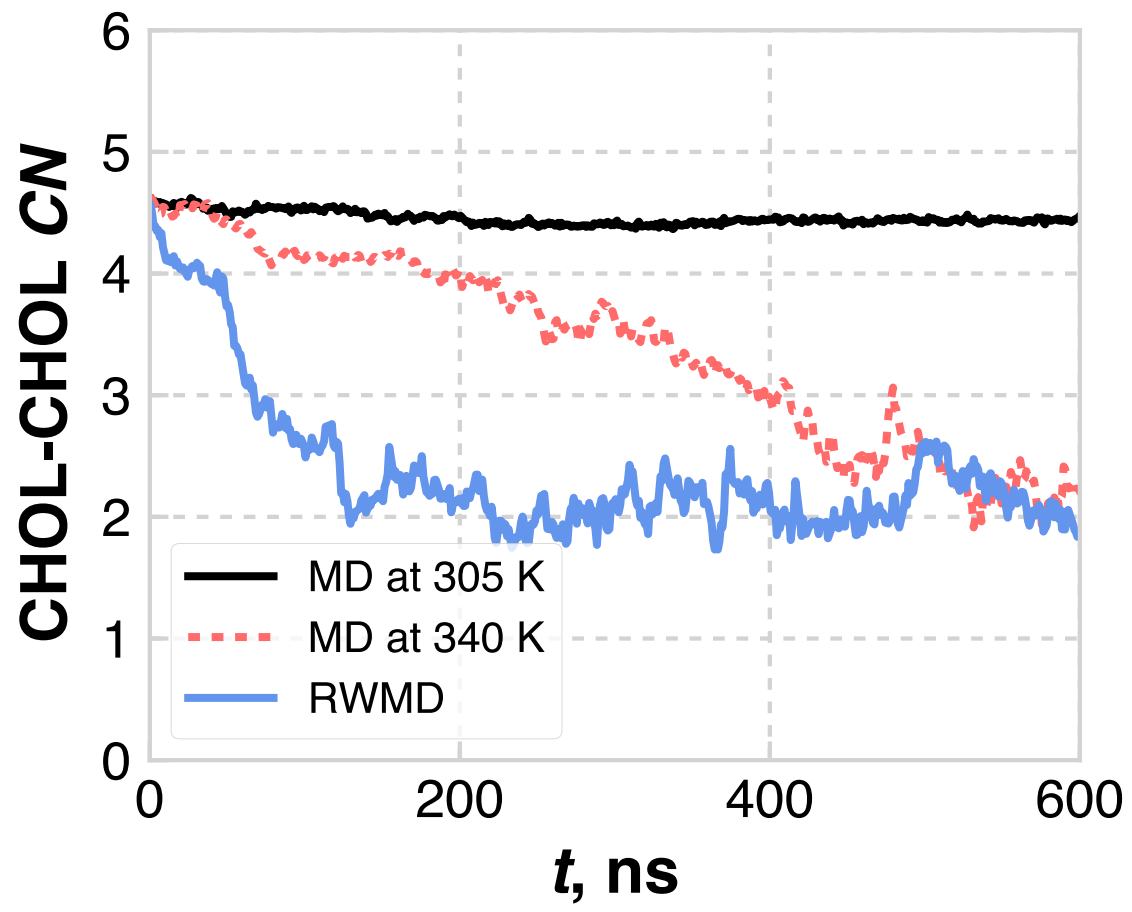

**Figure 4.** Evolution of the CHOL-CHOL coordination number (*CN*) for the fully phase-separated systems equilibrated with three different methodologies as a function of time (*t*).

**Dependence of Structural Properties on Bilayer Composition**

The effect of the composition of CER-based lipid bilayers on structural properties is now examined. Note, all systems start from a randomly dispersed morphology, given that, in the prior section, this arrangement relaxed the fastest and appears to be reasonably representative of the morphology for this family of systems; these systems were also relaxed using RWMD as described in Methods.

*General Structural Properties*. All systems exist in dense, highly ordered, gel phases, as indicated by high nematic order parameters ranging from 0.95 to 0.99, shown in Figure S4 and Table S1. The addition of CHOL decreases the nematic order, consistent with Das *et al*. (26), but the magnitude of this decrease is quite small, (e.g., ranging from 0.988 for pure uCER to 0.954 for 1-1 CER-CHOL with 3-1 eCER-uCER) and the lipids are still highly ordered from visual inspection (**Figure 1**e), indicating no phase change upon addition of CHOL for these compositions.

The tilt angle shows a weak dependence on composition, as shown in **Figure 5**a. The tilt angle depends significantly on the CER FA tail length for the pure CER systems, where a strong increase is seen with increasing eCER for systems with 50% or more eCER. This trend is a result of the balance between headgroup and tail interactions: steric repulsions between lipid headgroups dictate the *APL*, while van der Waals attractions between the lipid tails cause them to tilt to optimize their spacing (53). For the CER bilayers that are predominantly uCER, the tail length asymmetry gives rise to a wide *low density tail region* in the center of the bilayer (discussed below), which increases the effective optimal tail packing, leading to a lower tilt angle. The addition of CHOL and FFA causes the tails to tilt less and show almost no change as a function of increasing eCER, likely because CHOL has a bulky ring structure with a small headgroup, and therefore acts as a spacer between the lipid tails, buffering the head-tail effects seen in the pure systems. Hence, there is a smaller mismatch between the packing densities of the headgroups and tails, leading to a smaller tilt. Notably, the 1-1 CER-CHOL and ternary mixtures have very similar tilt angles, which we attribute to the fact that these systems have a similar fraction of alkyl tails.

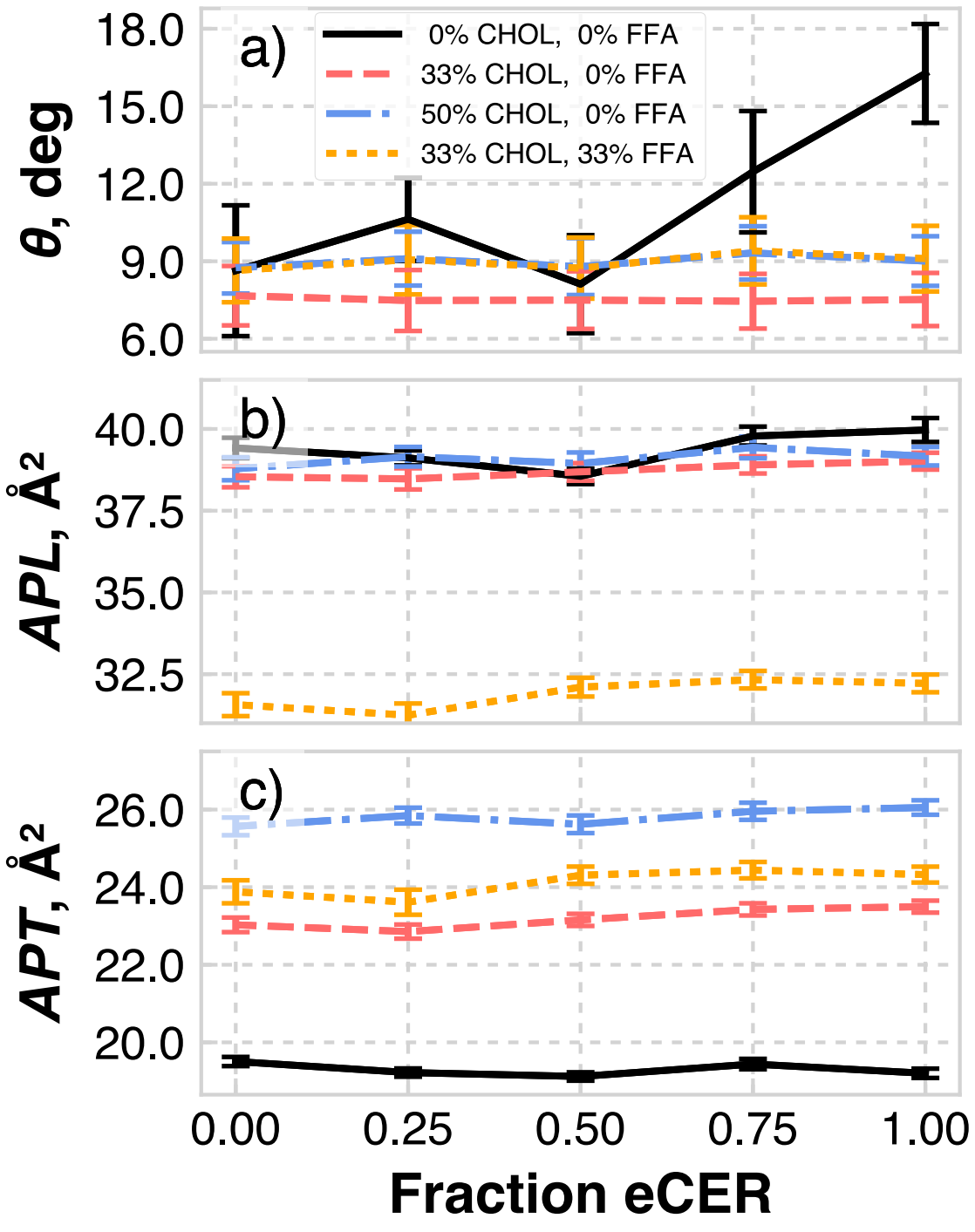

**Figure 5**. (a) Average tilt angle (*θ*) of the lipid tails with respect to the bilayer normal, (b) area per lipid (*APL*) and (c) area per tail (*APT*) as a function of eCER2 composition. Note that the abscissa is the fraction of CER NS that is eCER, not the total eCER fraction in the system.

The *APL* and *APT* trends as a function of the fraction of eCER are shown in **Figure 5**b,c. While the *APL* describes the density of lipids in the bilayer plane, irrespective of the number of tails that each lipid has, the *APT* describes the tail packing density in the plane of the lipid tails (i.e., *APT* accounts for both the number of tails and the tilt angle). All CER NS and CER NS-CHOL systems have similar *APL*s, between 38-40 Å$^2$, as expected since CER and CHOL have similar cross-sectional areas (54). There is only a slight, nonmonotonic dependence of the *APL* on the CER FA tail length. This trend is expected since the headgroups dictate the *APL* and these systems have the same headgroups for a given composition. Systems containing FFA have smaller *APL*s, since FFA contains a single alkyl chain. The *APT* increases with the CHOL content, as CHOL and CER NS have similar cross-sectional areas, but CHOL is treated as a single tail. The ternary systems have *APT*s more similar to the 2-1 CER NS-CHOL systems, which is reasonable considering they have the same ratio of CHOL to alkyl chains. The mixed lipid systems all show a slightly increasing *APT* with increasing eCER fraction. Interestingly, there appears to be some level of *APT* non-additivity in these systems. For example, consider the *APL* and *APT*s of the 1-1 CER NS-CHOL systems. Since they have similar *APL*s as the pure CER NS systems, but ¾ the number of tails, ideal mixing would yield *APL* values that are 4/3 that of the pure CER NS values. However, the actual values are lower, suggesting a relative attraction between CHOL and the saturated tails of CER NS, similar to what was observed for dimyristoylphosphocholine (DMPC, a phospholipid with saturated tails), but not dioleoylphosphocholine (DOPC, a phospholipid with unsaturated tails) (55). Although perhaps expected based on chain saturation, this behavior is nontrivial for CER systems, since the CER NS headgroups are much smaller than PC headgroups and CHOL presumably has less empty space to occupy.

*Density Profiles*. The bilayer thickness, calculated from the water density profiles, is a function of both overall lipid composition and CER FA tail length (Figure S5 and Table S2). The bilayer thickness decreases with increasing eCER and CHOL content, as one would expect since these lipids have shorter tail lengths than uCER or FFA. However, the tail length asymmetry leads to a richer behavior, which can be observed via examination of the lipid density profiles, as shown for each system in **Figure 6**.

The pure CER bilayers exhibit mass density profiles that contain the expected features of a lipid bilayer. Near the lipid-water interface, the profiles contain a peak (**Figure 6**a), which represents the heavy atoms in the CER headgroups. Just inside the headgroup region, there is a high density tail region where the tails are densely packed and highly ordered (see discussion of nematic order above), consistent with prior work (37). In the center of the bilayer is a low density tail region, where the lipid tails are less densely packed and less ordered. This region exists primarily due to the asymmetry in the lengths of the tails in the systems and is composed of the terminal part of the uCER tail and FFA tails, as shown in Figure S6. In the systems with less asymmetry (i.e., high eCER and CHOL concentrations), the profile is more V-shaped in the middle since there is little interdigitation between tails in opposing leaflets. Comparing the high and low density tail regions for the different bilayers, the width of the high density tail region is found to be nearly constant across all compositions, whereas the width of the low density tail region varies with eCER and CHOL content. **Figure 7** shows the total bilayer thickness as a function of the thickness of this low density tail region in the center of the bilayer. Clearly, the thickness of the low density tail region dictates the total bilayer thickness for a given composition. The width of the high density tail region is determined by the smallest hydrophobic length in the system, where the tails must pack densely to fit in the area dictated by the headgroups. For the lipids in this study, this is roughly 16 carbons, or the length of the sphingosine chain that is not part of the CER NS headgroup, which is also similar to the length of a CHOL molecule (56).

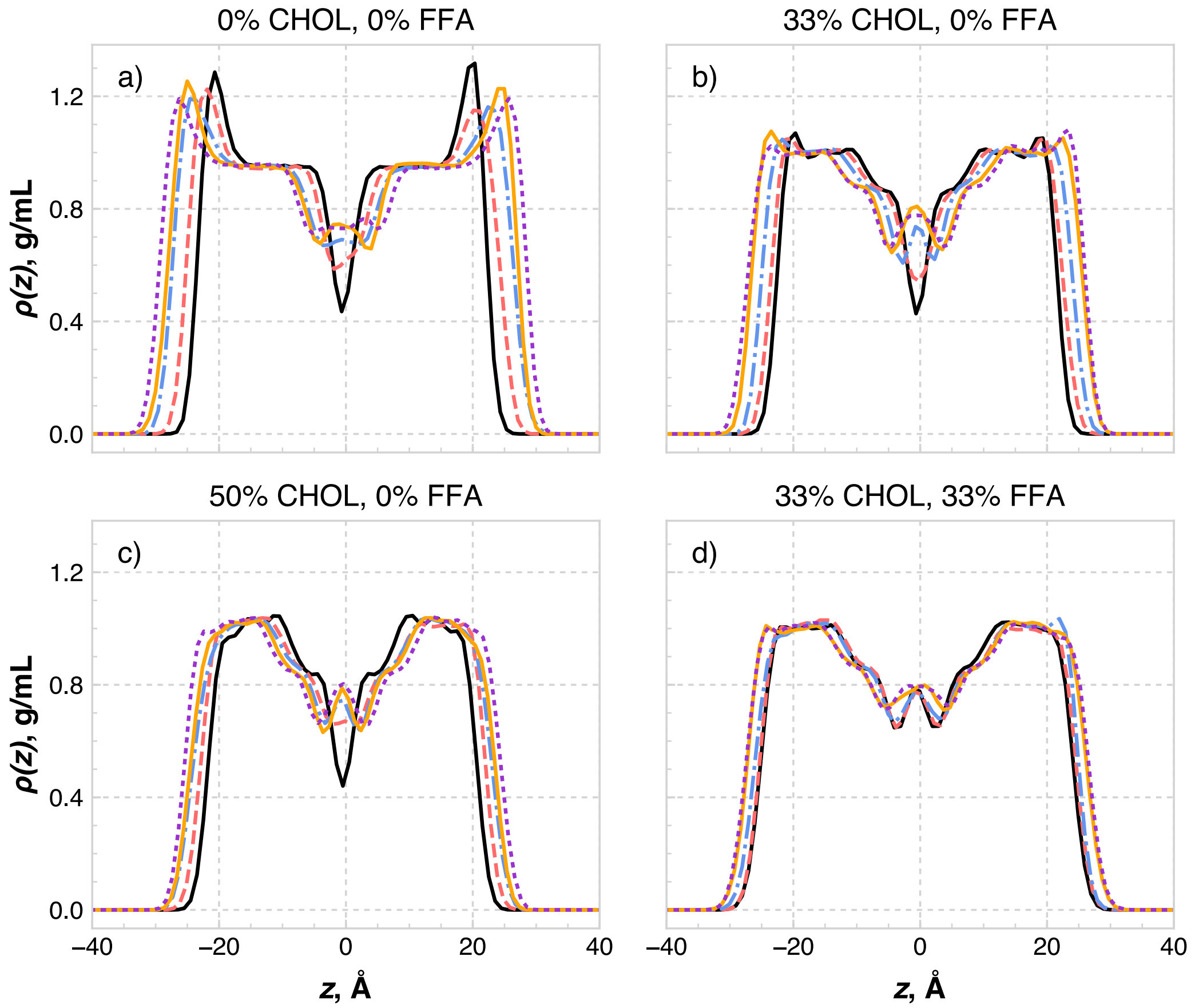

**Figure 6.** Total lipid mass density profiles across the bilayer for all systems considered: a) pure CER NS, b) 2-1 CER NS-CHOL, c) 1-1 CER NS-CHOL, d) equimolar CER NS-CHOL-FFA. The profiles for the 5 subcompositions are shown for all compositions and are labelled as follows: Solid black line 100 mol% eCER, dashed red line 75 mol% eCER, dot-dashed blue line 50 mol% eCER, solid orange line 25 mol% eCER, dotted purple line 0 mol% eCER. Note that these percentages denote the fraction of CER NS that is eCER, not the total fraction of eCER in the system.

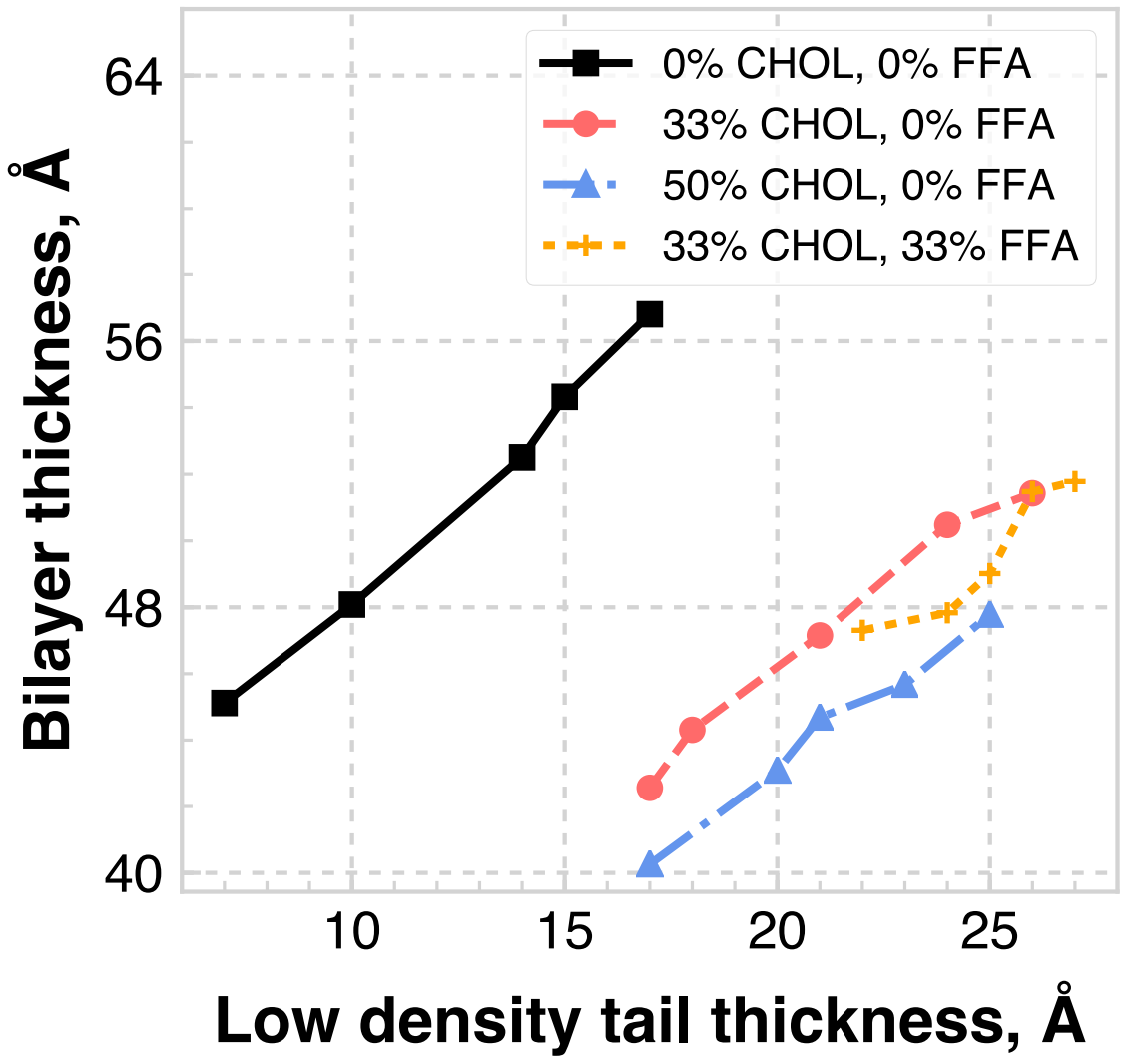

**Figure 7.** Total bilayer thickness as a function of the thickness of the low density tail region, with the different points on each line denoting the different eCER fractions; the smallest values indicate 100 mol% eCER, and the largest 0 mol% eCER.

The presence of CHOL and FFA has two main effects on the density profiles. First, the smaller headgroups of CHOL and FFA lead to smaller peaks in the headgroup region, compared to a pure CER NS bilayer; these peaks are dramatically reduced going from pure CER to 2-1 CER NS-CHOL and are mostly absent in the 1-1 CER NS-CHOL and ternary systems. Second, the width of the low density tail region increases with the addition of CHOL and FFA, as shown in **Figure 6** and **Figure 7**. Additionally, the low density tail regions tend to have more features in the mixed lipid systems, e.g., shoulders representing intermediate densities due to the different tail lengths. Peaks also appear in the middle of the density profiles for the mixed systems, due both to the long FA chain of uCER (and FFA for the 3 component system) and the presence of CHOL that has migrated from the ordered bilayer region into the low density tail region (28). In the eCER-CHOL systems, the lack of significant interdigitation means this peak is absent;

however, all systems containing FFA demonstrate a peak in the middle of the density profile because of the long tail of FFA.

**Figure 8** shows the density profiles for the ternary system with 1-1 eCER-uCER, along the bilayer normal direction, broken down by the contribution of specific lipid components. The CHOL headgroup sits ~3 Å deeper into the bilayer with respect to the CER headgroups, which is consistent with prior simulation and experimental studies (23, 26, 31). There is also a small subpopulation of CHOL lying flat in the middle of the bilayer, illustrated by the peak in the middle of the CHOL ring density profile, which is consistent with the work of Das *et al*. (28) Additionally, the FFA headgroups tend to sit slightly further into the water than the CER headgroups.

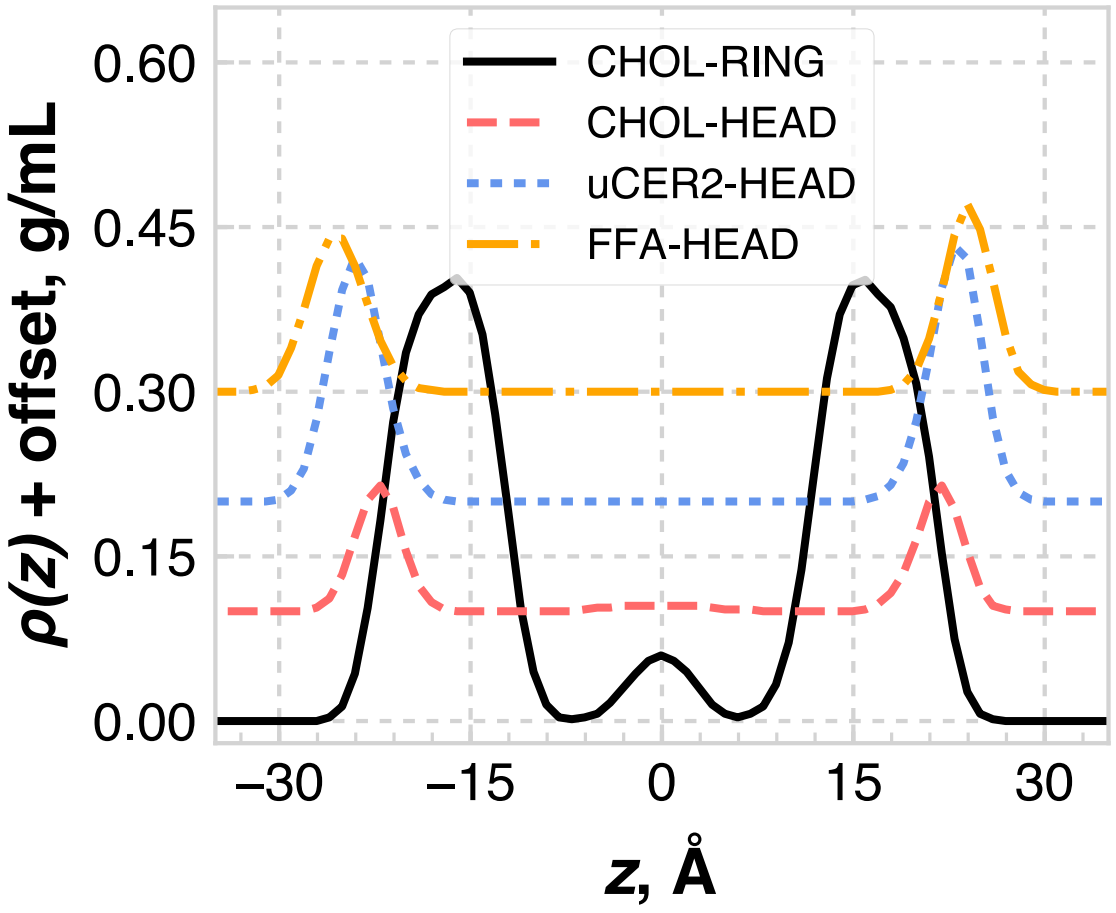

**Figure 8.** Mass density profiles of various groups in the ternary bilayer with a 1-1 eCER2-uCER2 composition. Note that the profiles for different groups are shifted vertically for clarity. Each curve is offset by 0.1 g/mL for each subsequent dataset in the legend.

The interfacial width, which can indicate the hydrophobicity of the surface (e.g., a larger interfacial width would indicate a more hydrophilic surface), is defined as the width of the region over which the water density drops from the bulk value to 1/*e* of the bulk value. The bilayers in this work appear hydrophobic, having interfacial thicknesses between 3.5-6 Å$^2$, compared to phospholipid bilayers, which are generally greater than 10 Å (57). There is little dependence on the bilayer composition or CER FA tail length, as shown in Figure S7 and Table S3. Thus, we conclude that while the lamellar organization near the middle of the bilayer depends on the lipid composition, the apparent hydrophobicity of the lipid-water interface does not.

*Hydrogen Bonding*. The total number of lipid-lipid hydrogen bonds in each system is listed in **Table 3**. The lipid-lipid hydrogen bonding is a function of the total CER content, with a reduction in CERs resulting in fewer lipid-lipid hydrogen bonds because CHOL and FFA molecules have fewer hydrogen bonding sites. The lipid-lipid hydrogen bonding is also independent of the eCER fraction, which may be expected since eCER and uCER have identical headgroups and their relative position at the interface is essentially unchanged as a function of the eCER2 fraction.

**Table 3.** Total number of lipid-lipid and lipid-water hydrogen bonds for each system studied.

| | | Fraction eCER | | | | |
|---|---|---|---|---|---|---|
| | **Composition** | **0** | **0.25** | **0.50** | **0.75** | **1.0** |
| **Lipid-lipid** | *Pure CER NS* | 63 | 62 | 61 | 63 | 63 |
| | *2-1 CER NS-CHOL* | 40 | 37 | 38 | 39 | 38 |
| | *1-1 CER NS-CHOL* | 28 | 26 | 28 | 28 | 27 |
| | *Ternary* | 19 | 19 | 19 | 19 | 19 |
| **Lipid-water** | *Pure CER NS* | 173 | 168 | 173 | 176 | 171 |
| | *2-1 CER NS-CHOL* | 173 | 169 | 173 | 175 | 176 |
| | *1-1 CER NS-CHOL* | 168 | 170 | 164 | 167 | 173 |
| | *Ternary* | 154 | 152 | 158 | 158 | 155 |

The hydrogen bonding between specific pairs of lipids for the ternary mixture with 1-1 eCER-uCER is listed in Table S4. Negligible CHOL-CHOL hydrogen bonding is observed, which can be rationalized by the fact that CHOL only has 1 hydrogen bond donor and acceptor, and two neighboring CHOL molecules would have to adopt a strained configuration to form a hydrogen bond. There are also very few CHOL-FFA hydrogen bonds, likely a result of CHOL sitting deeper into the bilayer. An appreciable number of CER-CHOL hydrogen bonds are seen, with slightly more eCER-CHOL (4.3) than uCER-CHOL (3.7) hydrogen bonds. Interestingly, the level of CER-CHOL hydrogen bonding is similar in the 2-1 and 1-1 CER-CHOL systems with, for example, 11.4 and 11.1 total CER-CHOL hydrogen bonds for the 2-1 and 1-1 CER-CHOL systems with 50 mol% eCER (Table S5 and Table S6); this trend may suggest that the CER-CHOL hydrogen bonding is saturated at or below 33 mol% CHOL. Additionally, when CER is replaced by CHOL (i.e., comparing the pure CER NS and the 2-1 CER NS-CHOL or 1-1 CER NS-CHOL systems), there are fewer hydrogen bonds relative to the total number of hydrogen bonding sites in the system, indicating that CER-CER hydrogen bonds are preferred over CER-CHOL hydrogen bonds. There are comparatively fewer CER-CHOL hydrogen bonds in the ternary systems, since FFA is present and is competing with CHOL to form hydrogen bonds with the CER. There is a surprisingly low amount of FFA-FFA hydrogen bonding, given that FFA prefers to be near other FFA molecules (discussed below); however, FFA sits deeper into the water and thus forms more hydrogen bonds with water.

The amount of lipid-water hydrogen bonding in each system is also listed in **Table 3**. Despite the relatively hydrophobic lipid-water interface, there are significantly more lipid-water hydrogen bonds than lipid-lipid hydrogen bonds. As with the lipid-lipid hydrogen bonds, the lipid-water hydrogen bonding is independent of the CER FA tail length. Interestingly, the pure CER and 2-1 CER-CHOL systems have similar amounts of lipid-water hydrogen bonds, despite the fact that CHOL has significantly fewer hydrogen bonding sites than CER, while the 1-1 CER-CHOL and ternary systems have comparably less lipid-water hydrogen bonds, since these systems have less CER (and hence fewer hydrogen bonding sites) than the systems with other two compositions. This trend in lipid-water hydrogen bonding is perhaps explained by the "spacer" effect of CHOL, discussed above with respect to the tilt angles. Although less CER results in fewer available hydrogen bonding sites, the spacing effect of CHOL allows the water to more thoroughly hydrate the CER headgroups and hence results in more lipid-water hydrogen bonds *per hydrogen bonding site*.

*In-Plane Morphology*. For a general view of the in-plane morphology of each system, we examine coordination numbers (CNs) between specific lipid tails in the bilayer plane. The CHOL-CHOL CNs depend on the amount of CHOL and available neighbors in a given system, and not on the CER FA tail length, as detailed in Figure S8 and Table S7. This result suggests a random distribution of CHOL throughout the bilayer leaflets. For example, there are equal numbers of three different types of tails in the 1-1 CER-CHOL systems (i.e., CER-FA, CER-SPH, and CHOL). Random mixing would give the observed CHOL-CHOL CN of 2. Applying this logic to the 2-1 CER-CHOL and ternary systems, random mixing would imply CHOL-CHOL CNs of 1.2 and 1.5 for each composition, respectively, which is observed. In contrast to CHOL, FFA shows a preference for specific neighbors, and thus has a less random distribution throughout the bilayer leaflets. In the uCER-CHOL-FFA system, the FFA-$CER_{FA}$ and FFA-$CER_{SPH}$ CNs are 1.39 ± 0.08 and 1.20 ± 0.07, respectively (where $CER_{FA}$ and $CER_{SPH}$ represent the fatty acid and sphingosine tails of CER, respectively). This result suggests that FFA has a preference for the FA chain of CER compared to the SPH chain. Additionally, a significant preference for FFA to be near FFA is observed, with a FFA-FFA CN of 2.3 ± 0.1, compared to 1.39 ± 0.08 for FFA-$CER_{FA}$, the second highest. This has also been observed experimentally, with FFA-enriched domains forming with the tails tightly packed on an orthorhombic lattice (19). However, unlike experimental results on similar systems (41), we do not see any chain length-dependent behavior in regards to the mixing of CER and FFA. We note though that comparable experimental systems consider multilamellar structures, whereas we consider a single hydrated bilayer, which could account for the differences observed. We also note that the systems studied in the current work are several orders of magnitude smaller than comparable experimental systems (nm vs. $\mu$m), and that an in-depth study of lateral distributions of lipids is beyond the scope of this work.

## DISCUSSION

*Relevance to Experimental Models of the Short Periodicity Phase*. The localization of different groups within an experimental model of the short periodicity phase (SPP) of the SC has been studied with neutron diffraction (22, 23). These experimental systems resemble the systems studied here, with a few differences. First, we only consider CER NS, whereas experimental systems tend to include a mixture of CERs with different headgroups, although uCER typically accounts for the majority (~60%) of the CERs (11, 23). Secondly, we also only consider FFA C24, whereas experimental systems often include FFAs with a distribution of tail lengths. We note, however, recent work has shown that CER headgroup chemistry does not play a strong role in the lamellar organization of SPP models, and that systems with a distribution of FFA tail lengths had the same lamellar organization as a similar system with only FFA C24 (11). Perhaps the biggest difference between experimental systems and the systems studied here is that we consider a single bilayer, which is the typical approach to studying these systems with molecular simulation. Experimental systems usually contain membranes with numerous diffraction orders detected, indicating multilayer structures. Nonetheless, we can directly compare neutron scattering length density (*NSLD*) profiles from simulation and experiment. Additionally, we can compare the localization of specific groups based on mass density profiles from simulation and those reconstructed from selective deuteration of specific groups from experiment.

In **Figure 9** we compare the total *NSLD* profiles from experiment (22, 23) and simulation. The most obvious features of each are the large peaks at ±27 Å (experiment) and ±25 Å (simulation). Given the experimental resolution of ~5.4 Å, the locations of these peaks are in good agreement. Both simulation and experiment show valleys in the *NSLD* at ~7 Å, which we note aligns with the dips in the low tail density region from **Figure 6**d. Additionally, subtle shoulders in the *NSLD*s are present at ±15 Å in both simulation and experiment, which aligns with the edges of the low tail density region shown in **Figure 6**d.

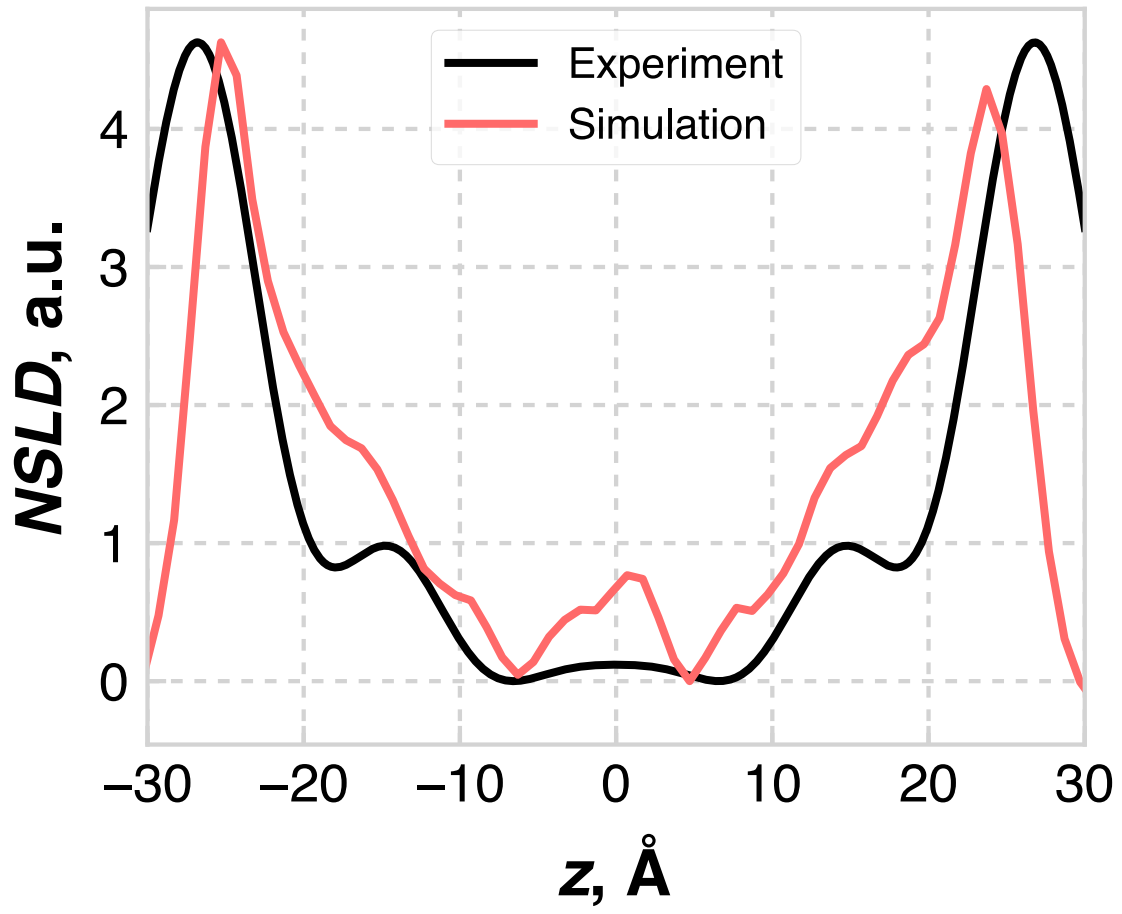

**Figure 9.** Neutron scattering length density (*NSLD*) profiles, reconstructed from experiment(22) and calculated from simulation. Note that the profiles were shifted such that the minimum lies at 0, and the simulation curve was scaled to have the same height as the experimental curve.

Focusing next on the localization of specific groups within the bilayer, we compare the mass density profiles from simulation with the difference between the protonated and deuterated *NSLD*s from experiment, shown in Figure S9. In this manner, we can compare the localization of specific groups between simulation and experiment. Comparing the location of the FA tail of uCER that was measured in Groen *et al.*, (22) we find good agreement, with a single broad peak spanning ±7.5 Å (Figure S9a), which corresponds to the region where the uCER FA tails from opposing leaflets interdigitate. The CHOL tails are localized at ±2 Å in both simulation and experiment (Figure S9b) (23). The CHOL headgroups show some deviation between experiment and simulation, localized at ±23 Å and ±21 Å, respectively (Figure S9c). This discrepancy, however, is again within the experimental resolution of 5.4 Å (23). Despite the fact that we are studying single bilayers, the locations of specific groups within the bilayer agree very well with experimental data on systems with similar compositions. Therefore, we can conclude that the model systems accurately approximate the model SPP systems from experiment, and conclusions from this work also likely apply to model SPP systems, and also the SPP in the SC.

*Effects on Barrier Properties of SC*. In this work, the lipid composition and CER FA tail length were found to have the largest impact on lamellar organization, with the bilayer thickness and shape of the density profile in the low density tail region most affected. If we consider how these changes would affect the barrier properties (i.e., the permeability) of these bilayers, we can provide some insight into the observed experimental behavior. The permeability of a membrane is a product of the solute diffusion and partitioning, where the partitioning has an exponential dependence on the free energy of solvation. Due to this exponential dependence, small changes in local packing (i.e., depth-dependent density), can have a dominant impact on the bilayer permeability. Since the lipid mass density profiles are not constant across the bilayers, the resistance experienced by a permeant molecule would vary with bilayer depth. Moreover, since the low density tail regions are qualitatively different for the bilayers with different compositions, these depth-dependent partition coefficients would also change, albeit nontrivially. Thus, if one was interested in comparing the permeability of these different systems, a simple homogeneous solubility-diffusion model would not suffice. This indeed is not unrecognized in the field, as the few simulation studies of SC lipid permeability in the literature to date have used an inhomogeneous solubility-diffusion model (30, 52).

While the structure of the lipids at the lipid-water interface changes with composition, it does not change with CER FA tail length. This is illustrated by the fact that the thickness of the interfacial region and lipid density profiles near the headgroups are unchanged with the CER FA tail length for a given composition (**Figure 6** and Figure S7). Additionally, both the total number of lipid-water and lipid-lipid hydrogen bonds is constant for a given composition. Thus, we expect any changes in the barrier properties of model SC membranes with CER FA tail length to be a result of changes in the structure of the bilayers in the hydrophobic core, and that changes in the headgroup region would play only a small role. This observation is consistent with the permeability measurements reported by Uchiyama *et al.*, which show that the permeability of ethyl-*p*-aminobenzoic acid through membranes composed of synthetic mixtures of CER, CHOL, and FFA is independent of the CER composition, but strongly dependent on the FFA tail length dispersity (11).

## CONCLUSIONS

We have studied via molecular dynamics simulations the structure of lipid bilayers relevant to the stratum corneum layer of the skin, as well as proposed and validated a thorough relaxation protocol for such systems. We first illustrated several difficulties in simulating SC lipid mixtures and multicomponent gel-phase bilayers in general e.g., systems tend to be frozen at physiological temperatures with only small lateral rearrangements possible. Since the most realistic morphologies are not generally known *a priori*, systems must be allowed to find the most realistic morphology. This task is shown to be computationally impractical at physiological temperatures and inefficient at commonly used elevated temperatures, as well as likely to lead to metastable states being confused for equilibrium states. The proposed random walk MD methodology, based on simulated tempering, was shown to be an efficient and reproducible method for minimizing the influence of an assumed, initial membrane configuration for multicomponent, gel-phase bilayers. We therefore expect this method to be especially useful as simulations incorporate more complex lipid mixtures. Using the random walk MD protocol, we examined a series of SC lipid bilayers with varying lipid compositions and CER FA tail length. We showed that the lamellar organization is most affected by the lipid composition and CER FA tail length. Subtle changes to the lateral organization of the lipid tails as a function of the CER FA tail length were also seen. Additionally, since it was observed that the behavior at the lipid-water interface does not change with the CER FA tail length, we speculate that any tail length-dependent changes in barrier function are a result of changes in the lipid tail region and not the headgroup region.

**Supporting Information**. The Supporting Information contains the content listed below.

An explanation of the various metrics used to quantify the structure of the bilayers in this study.

Figure S1. Representative plot of temperature versus time during RWMD equilibration.

Figure S2. Schematic of the bilayer thickness and interfacial thickness calculations.

Figure S3. Visualization of the low tail density region.

Figure S4. Nematic order parameter of lipid tails.

Figure S5. Bilayer thickness as a function of lipid composition.

Figure S6. Mass density profiles of the "unequal" part of the uCER2 tail and terminal 8 carbons of the FFA tail in the equimolar uCER2-CHOL-FFA system.

Figure S7. Thickness of lipid water interface as a function of system composition.

Figure S8. CHOL-CHOL coordination number as a function of system composition.

Figure S9. Localization of various lipid components compared to neutron scattering results.

Table S1. Nematic order parameter of lipid tails as a function of bilayer composition.

Table S2. Bilayer thickness as a function of bilayer composition.

Table S3. Thickness of lipid-water interface as a function of bilayer composition.

Table S4. Hydrogen bonding, broken down by lipid component, for the ternary system with a 1-1 eCER2-uCER2 ratio.

Table S5. Hydrogen bonding, broken down by lipid component, for the 2-1 CER2-CHOL system with a 1-1 eCER2-uCER2 ratio.

Table S6. Hydrogen bonding, broken down by lipid component, for the 1-1 CER2-CHOL system with a 1-1 eCER2-uCER2 ratio.

Table S7. CHOL-CHOL coordination numbers as a function of system composition.

## AUTHOR CONTRIBUTIONS

C.M., C.R.I., and A.L.B. designed research; T.C.M., and C.R.I. performed research; T.C.M. and R.H. contributed analytic tools; all authors analyzed data and wrote the article.

## ACKNOWLEDGEMENTS

This work was supported by grant number R01 AR057886-01 from the National Institute of Arthritis and Musculoskeletal and Skin Diseases and National Science Foundation grant number CBET-1028374. This work was conducted in part using computational resources provided by the National Energy Research Scientific Computing Center, supported by the Office of Science of the Department of Energy under Contract No. DE-AC02-05CH11231 and the Advanced Computing Center for Research and Education at Vanderbilt University.